\documentclass{article}%
\usepackage{amssymb}
\usepackage{graphics}%
\usepackage{amsmath}%
\usepackage{amsfonts}%
\usepackage{graphicx}
\begin{document}

\title{Continuum percolation and the liquid--solid coexistence line of simple fluids.}
\author{Luis A. Pugnaloni\thanks{E-mail: luis@iflysib.unlp.edu.ar}, Marcos G. Valluzzi
\and and Fernando Vericat\\Instituto de F\'{\i}sica de L\'{\i}quidos y Sistemas Biol\'{\i}gicos,\\UNLP-CONICET - cc. 565, 1900 La Plata, Argentina}
\maketitle

\begin{abstract}
We compare the percolation loci for chemical clusters with the liquid--solid
transition in the temperature--density phase diagram. Chemical clusters are
defined as sets of particles connected through particle-particle bonds that
last for a given time $\tau$. By using molecular dynamics simulations of a
Lennard--Jones system we obtain the percolation loci at different values of
$\tau$ as the lines in the temperature--density plane at which the system
presents a spanning cluster in 50 percent of the configurations. We find that
the percolation loci for chemical clusters shifts rapidly towards high
densities as $\tau$ is increased. For moderate values of $\tau$ this line
coincides with the low-density branch of the liquid--solid coexistence curve.
This implies that no stable chemical clusters can be found in the fluid phase.
In contrast, the percolation loci for physical clusters ---sets of particles
that remain close together at every instant for a given period $\tau$--- tends
to a limiting line, as $\tau$ tends to infinity, which is far from the
liquid--solid transition line.

\end{abstract}

\section{Introduction}

Clustering and percolation in continuum systems are concepts of great interest
in many areas of physics. See for example ref.
\cite{Stauffer,Senger1,Starr1,Chen1,Butler1,Pugnaloni0,Pugnaloni0b}. In a
many-body disordered system, the central problem in continuum percolation
consists in determining the critical density (for a given temperature) at
which the system presents a cluster that spans the system volume. This, in
turn, leads to the crucial question of what is a cluster. The answer to this
question is many-sided. A cluster is a set of particles that can be considered
as a unit separated from the rest of the system and that can manifest itself
in an experiment. What we define as a cluster in a given system depends on the
properties of interest. Therefore, there is a degree of arbitrariness in the
cluster definition. The main premise here is that any proposed definition
allows a separation of the system into a number of disjoint sets of particles
which is appropriate to describe the phenomenon under study. For example, if
one intends to explain an insulator--conductor transition, and the particles
of the system are the conducting objects, we should call cluster to any set of
contacting particles that is not in contact with any other particle in the
system \cite{Chen1}. A good example is also found in deposited granular matter
where clusters correspond to mutually stabilizing sets of particles called
arches \cite{Pugnaloni0,Pugnaloni0b}.

The first physically sound cluster definition for a simple fluid was given by
Hill \cite{Hill1} by contrast with the mathematical clusters introduced by
Mayer \cite{Mayer1}. In Hill's theory, the concept of cluster is directly
related to the idea of \textit{bonded pairs}. A bonded pair is a set of two
particles that are linked by some direct mechanism. Then, a cluster is defined
as a set of particles such that any pair of particles in the set is connected
through a path of bonded pairs. The definition of bonded pair becomes then the
essential part of the theory and a number of definitions have been used. Among
others, we can mention the energetic criterion and the geometric criterion for
bonded pairs proposed by Hill himself \cite{Hill1}. In the first case, two
particles are considered bonded at a given time if the negative of the pair
potential energy is higher than the relative kinetic energy of the pair (see
for example refs. \cite{Campi1} and \cite{Pugnaloni4}). In the second case,
two particles are bonded at a given time if they are within a distance $d$.
These later geometrical clusters are referred to as Stillinger's clusters (see
for example refs. \cite{Stillinger1} and \cite{Heyes1}).

In a previous paper \cite{Pugnaloni1}, we have introduced a new definition of
cluster based on Hill's theory where the bonded pairs are defined as particles
that remain within a certain distance $d$ during a time interval $\tau$. We
call these clusters \textit{chemical clusters}\ since the stability
requirement on each connection remind us of a covalent bond. Note that a very
similar dynamic definition was advanced previously by Bahadur and McClurg
\cite{Bahadur1}. In the same paper, we have also presented a different
approach to cluster identification where the concept of bonded pairs is
meaningless. In this case, we identify as a cluster any set of particles that
remain interconnected (in the Stillinger's sense) at every instant over a time
interval $\tau$. Each particle within a given cluster may move around without
forming a stable bond with any other particle whereas it stays connected to
the group. We have termed these later clusters \textit{physical clusters} for
they exist without the presence of specific particle--particle bonds.

Some preliminary studies of the properties of chemical and physical clusters
in Lennard--Jones systems where presented in ref. \cite{Pugnaloni1}. Moreover,
an integral equation theory was presented there ---and later on successfully
used \cite{Zarragoicoechea1}--- to calculate the cluster pair correlation
function for chemical clusters. Also, a first theoretical attempt to treat the
case of physical clusters was advanced in ref. \cite{Pugnaloni1}. In this
paper, we present the results of an exhaustive investigation of the
percolation of chemical clusters for a variety of values of $\tau$. We have
encountered that chemical clusters percolate at progressively higher densities
as $\tau$ grows from zero. Beyond a moderate value of $\tau$, the percolation
line coincides with the low-density branch of the liquid--solid coexistence
curve. This represents, to our knowledge, the first application of continuum
percolation where the percolation line coincides to such degree with a
coexistence line of a simple fluid. We also show that the percolation line for
physical clusters, although it moves towards higher densities as $\tau$ grows,
reaches a maximum density within the fluid phase beyond which the system
percolates irrespective of how large is $\tau$. From these findings, we
conclude that stable chemical clusters cannot exist in a simple fluid although
physical cluster do exist at relatively high densities. We also point out that
transient chemical clusters (by contrast with stable clusters) may help to
explain differences in experiments carried out on the same system when probing
different timescales.

\section{Stillinger's clusters}

In a many-body system, two particles at time $t$ are considered to form a
\textit{bonded pair} (in the Stillinger's sense) of proximity $d$, if they are
within a distance $d$ from each other at time $t$. Then, a set of particles at
time $t$ is called a \textit{Stillinger's cluster} of proximity $d$, if every
particle in the set is connected through a path of pair-bonded particles and
no other particle in the system is bonded to any of them. This cluster
definition is a basic standpoint for the following more realistic definitions.

\section{Chemical clusters}

Two particles at time $t$ are considered to form a \textit{bonded pair} of
lifespan $\tau$ and proximity $d$, if they were within a distance $d$ from
each other over the entire time interval $[t-\tau,t]$. Then, a set of
particles at time $t$ is called a \textit{chemical cluster} of lifespan $\tau$
and proximity $d$, if every particle in the set is connected through a path of
pair-bonded particles and no other particle in the system is bonded to any of
them. A chemical cluster with lifespan $\tau\gg\tau_{0}$ is called a
\textit{stable} chemical cluster. Here, $\tau_{0}$ means any time long enough
for the cluster to be detected by an experimental setup. A chemical cluster of
lifespan $\tau=0$ and proximity $d$ corresponds to a Stillinger's cluster of
proximity $d$.

We would not normally expect to find stable chemical clusters in a simple
fluid since particles do not tend to form stable particle--particle bonds.
However, if we use an experimental technique able to resolve short times, we
might be able to detect the presence of ephemeral (small $\tau$) chemical
clusters. Moreover, if interactions are significantly stronger and dynamics
significantly slower, as is the case in colloidal systems, most experiments
are able to detect the existence of chemical clusters (see for example the
relation between percolation of chemical clusters and the sol-gel transition
\cite{Pugnaloni3}). Of course, chemical clusters are easily detected in true
chemical bonding (\textit{e.g.} polymerisation) and in strongly interacting
molecules (\textit{e.g.} hydrogen bonding \cite{Sciortino1}).

\section{Physical clusters}

Suppose that we identify a given set of particles at time $(t-\tau)$ as a
Stillinger's cluster of proximity $d$. We then focus on this set of particles
alone. As the system evolves in time, the particles of this Stillinger's
cluster approach other particles in the system, which we will disregard in our
analysis. Eventually, the original Stillinger's cluster fragments into two or
more disconnected subclusters --- in the Stillinger's sense. Each of these
subclusters can be now considered individually, disregarding the remaining
particles in the system --- even those that were part of the parent cluster
and now belong to a different subcluster. Again, during the evolution, each
subcluster may fragment. If we continue this analysis systematically until
time $t$ is reached (\textit{i.e.} over a period of time $\tau$), each final
subcluster, which is a product of numerous fragmentation events, is called a
\textit{physical cluster} of lifespan $\tau$ and proximity $d$. A physical
cluster with lifespan $\tau\gg\tau_{0}$ is called a \textit{stable} physical
cluster. Again, $\tau_{0}$ represents a time long enough for the cluster to be
detected by an experimental setup. A physical cluster of lifespan $\tau=0$ and
proximity $d $ corresponds to a Stillinger's cluster of proximity $d$.

We expect to find stable physical clusters in a simple fluid. In fact, any
sample of liquid should correspond to a single physical cluster.

\section{Molecular dynamics}

The molecular dynamics (MD) simulation technique used to identify both
chemical and physical clusters was introduced and described in ref.
\cite{Pugnaloni1}. This technique is a slight modification of a standard $NVT$
MD simulation \cite{Allen1} where bonded pairs (for chemical cluster) and
entire Stillinger's clusters (for physical clusters) are tracked down along a
time interval of length $\tau$. We report results for a Lennard--Jones system,
\textit{i.e.} particles that interact through%
\begin{equation}
v(r)=4\varepsilon\left[  \left(  \frac{\sigma}{r}\right)  ^{12}-\left(
\frac{\sigma}{r}\right)  ^{6}\right]  .\label{e1}%
\end{equation}
Quantities are then expressed in Lennard--Jones reduced units in the rest of
the paper: $\sigma$ for length, $\sigma\sqrt{m/\varepsilon}$ for time, and
$\varepsilon$ for energy. Quantities in reduced units are indicated with an asterisk

The results that we present in the following section are based on a leapfrog
MD simulation of $500$ particles in a cubic box with periodic boundary
conditions. A cut off distance equal to $2.7\sigma$ was used in the
Lennard--Jones pair potential. We have set the proximity value to $d^{\ast
}=1.5$. This value was chosen because it corresponds roughly to the maximum
distance at which two Lennard--Jones particles attract most strongly. Besides,
at $d^{\ast}=1.5$ lies the boundary between the first and the second neighbour
shell in the supercooled liquid state \cite{Keyes1}. This can be considered
the maximum distance for which a particle should be considered directly
connected from a geometrical point of view.

In most cases, quantities are averaged over $10^{3}$ configurations generated
after stabilization. A few simulations performed with 4000 particles yielded
results indistinguishable from those obtained with the 500-particle-system. A
system is said to be in a percolated state if a cluster that spans the
periodic replicas is present 50 percent of the time \cite{Seaton1}. Then, a
percolation transition curve (the percolation loci), which separates the
percolated and the non-percolated states of the system, can be drawn above the
gas--liquid coexistence curve in the $T-\rho$ phase diagram.

\begin{figure}[ptb]
\includegraphics{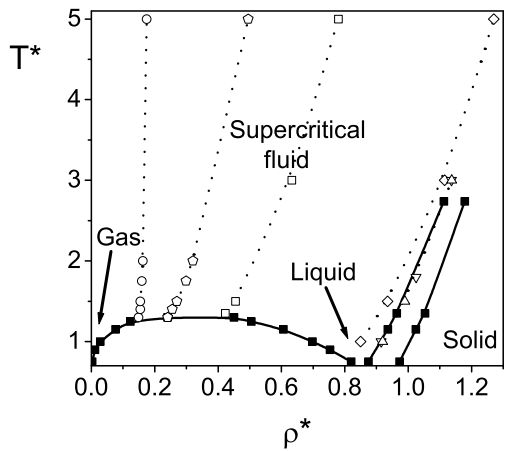}\caption{Temperature--density
phase diagram for the Lennard--Jones fluid and percolation loci for chemical
clusters. Solid squares (with trend line) correspond to the coexistence curve
for gas--liquid and liquid--solid transition. Open symbols (with dotted trend
line) correspond to the percolation loci for chemical clusters with proximity
$d^{\ast}=1.5$ and lifespan $\tau^{\ast}=$ 0 (circles), 0.5 (pentagons), 1.0
(squares), 5.0 (diamonds), 8.0 (down triangles), and 10.0 (up triangles).}%
\label{figure1}%
\end{figure}

\section{Results}

In fig. \ref{figure1} we show the phase diagram of the Lennard--Jones fluid in
the $T-\rho$ plane. The gas--liquid coexistence curve corresponds to the Gibbs
ensemble Monte Carlo (MC) simulation results from Panagiotopoulos
\cite{Panagio1} and the liquid--solid coexistence line corresponds to the MC
calculations of Hansen and Verlet \cite{Hansen1}. The percolation line for
chemical clusters of proximity $d^{\ast}=1.5$ and lifespan $\tau^{\ast
}=0,0.5,1,5,8$ and $10$ are also shown in fig. \ref{figure1}. The special case
with $\tau^{\ast}=0$ corresponds to \textit{instantaneous} Stillinger's
clusters. As we can see, the percolation line shifts towards high values of
$\rho$ as $\tau$ is increased. This is to be expected in a simple liquid
because, for a given temperature, the density needs to be increased in order
to promote caging and so extend the lifespan of particle--particle bonds.

\begin{figure}[ptb]
\includegraphics{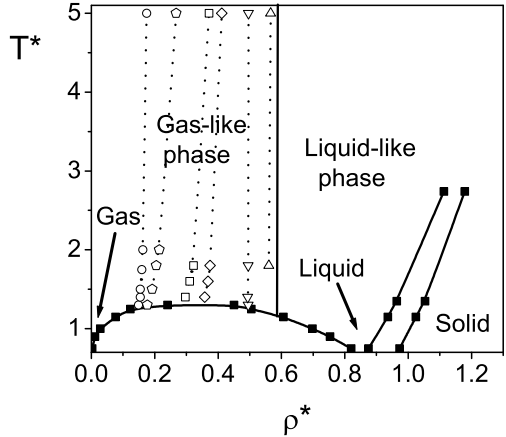}\caption{Temperature-density
phase diagram for the Lennard--Jones fluid and percolation loci for physical
clusters. Solid squares (with trend line) correspond to the coexistence curve
for gas--liquid and liquid--solid transition. Open symbols (with dotted trend
line) correspond to the percolation loci for physical clusters with proximity
$d^{\ast}=1.5$ and lifespan $\tau^{\ast}=$ 0 (circles), 1 (pentagons), 5
(squares), 10 (diamonds), 100 (down triangles), 1000 (up triangles). The solid
vertical line corresponds to the critical density $\rho_{f}$\ (see text).}%
\label{figure2}%
\end{figure}

Interestingly, for $\tau^{\ast}\gtrsim8.0$ we find that the percolation line
coincides with the liquid branch of the liquid--solid transition curve.
Rigorously speaking, if $\tau^{\ast}\gtrsim8$, any state of the system to the
left of the liquid--solid coexistence line is non-percolated whereas any state
in the solid side of the phase diagram is percolated. The reason for this
resides in the fact that simple liquids have to make a transition to a
crystalline state in order to generate long-lasting particle--particle bonds.
In a crystalline state the neighbours of each particle remain the same for
extremely long periods of time. Of course, a similar effect can be achieved in
a glassy state, but we are only concerned with equilibrium states here. From
this result we can say that stable ($\tau\gg\tau_{0}$) chemical cluster do not
exist in the fluid phase of a simple liquid. It is worth mentioning that using
different values of $d$ (at least in the range $1.25<d^{\ast}<$ $2.0$) shifts
all the percolation curves but they still converge to the liquid--solid
transition line for large $\tau$. Larger values of $d$ requires larger values
of $\tau$ to achieve convergence.

Our findings on the properties of the chemical cluster percolation line for
large $\tau$ suggests that we could use this analysis to predict the
liquid--solid transition line using a relatively simple simulation technique
and the appropriate cluster identification criterion. Previously, much effort
have been made to find a match between percolation and the gas--liquid
transition, especially in the two-state magnetic lattice model of a fluid
\cite{Stauffer2}. This studies have been unsuccessful so far, which can be
appreciated in the type of device (ghost spin) needed to prevent
inconsistencies in the theory \cite{Stauffer2}. As we can see, chemical
clusters are a more realistic definition of clusters and shows that the
percolation line should describe the liquid--solid transition rather than the
gas--liquid transition. Whether there exists a suitable definition of cluster
that yields a match between percolation and gas--liquid transition is still unresolved.

For the sake of completeness, the percolation behaviour of physical clusters
is shown in fig. \ref{figure2}. In contrast with chemical clusters, physical
clusters percolate at moderate densities (far from the liquid--solid
transition) even at very large values of $\tau$ ($\tau^{\ast}\simeq10^{3} $).
In fact, the percolation curve seems to converge to a limiting line beyond
which the system is always percolated no matter how large is $\tau$. This
suggest that there exist a critical density $\rho_{f}$ ($\rho_{f}^{\ast}%
\simeq0.573$) in the supercritical phase above which physical clusters
percolates irrespective of the temperature and the required lifespan
\cite{Pugnaloni5}. This line is sketched in fig. \ref{figure2}, and it can be
thought of as a dividing line between a gas-like phase (to the left of
$\rho_{f}$) and a liquid-like phase (to the right of $\rho_{f}$). Within the
gas-like phase, physical clusters percolate only at finite lifespan. In the
liquid-like phase physical clusters always percolate the system. Within the
liquid-like phase ---except in a small neighbourhood around $\rho_{f}$---
nearly all particles in the system remain interconnected in a single large
physical cluster, \textit{i.e.}, no fragmentation takes places within any
finite time interval.

The main objection we have to this separation into liquid-like and gas-like
regions of the supercritical phase is that states that we would normally
assign to a liquid reside in the gas-like side of the phase diagram. The
density $\rho_{f}$ is certainly higher than the thermodynamic critical density
$\rho_{c}$. All the states comprised between the gas--liquid coexistence
curve, the vertical line that passes through $\rho_{f}$ and the horizontal
line that passes through $T_{c}$ (the thermodynamic critical temperature) are
in a liquid state. According to our physical cluster properties, however, they
should be in a gas-like state. Interestingly, a different separation of the
supercritical phase proposed by Fisher and Widom \cite{Fisher1} does also meet
the gas--liquid coexistence line at a density above $\rho_{c}$ \cite{Vega1}.
However, the Fisher--Widom line does not coincide with the dividing line we
present here.

\section{Conclusions}

In this work, we have shown that chemical clusters and physical clusters
defined in ref. \cite{Pugnaloni1} have very different behaviour at large
lifespan. In particular, chemical clusters present a percolation curve that
coincides with the low-density branch of the liquid--solid transition curve
for any $\tau$ beyond a moderate value. Physical clusters, in contrast,
present a percolation curve that converges to a vertical line in the $T-\rho$
plane located at moderate densities when $\tau$ tends to infinity.

The percolation properties of chemical clusters suggest that we can predict
the liquid--solid transition in a simple fluid by calculating the percolation
line at moderate lifespan values. An integral equation theory for chemical
clusters has been developed previously \cite{Pugnaloni1,Pugnaloni2}. Moreover,
a numerical technique has been used recently \cite{Zarragoicoechea1}\ to solve
the equations, but only at small values of $\tau$. This technique may allow us
to predict the position of the liquid--solid coexistence line without the need
of free energy calculations by setting $\tau$ to larger values ($\tau^{\ast
}\gtrsim8.0$).

It is important to notice that is not possible to find stable (long lifespan)
chemical clusters in the fluid phase of a simple fluid. As soon as we exceed a
moderate value of $\tau$ we require the system to crystallise in order to find
chemical clusters of lifespan $\tau$. This finding seems rather intuitive;
however, until now, percolation models where not consistent with this picture.

For small values of $\tau$, we find a family of percolation curves for
chemical clusters ranging from the classical instantaneous (Stillinger's)
cluster percolation up to the liquid--solid transition. Each of these
percolations curves may be observable as viscoelastic sol--gel-like
transitions measured at different shear frequencies \cite{Pugnaloni3}. Indeed,
working at very high shear frequencies will result in a sol--gel transition
line at relatively low densities. High frequencies probe short timescales and
any ephemeral (small $\tau$) stress-bearing percolating cluster will develop a
dominating elastic response in the system. Conversely, working at very low
frequencies will result in a sol--gel transition line at relatively high
densities. Low frequencies probe long timescales and the particle--particle
bonds have to last long enough (large $\tau$) for any percolating cluster to
develop a dominating elastic response.

\textbf{Acknowledgments.}\ Support of this work by UNLP, CONICET and ANPCYT of
Argentina is very much appreciated. M.G.V. is a fellow of CONICET. L.A.P. and
F.V. are members of CONICET.

\end{document}